# Identification of nutrient deficiency in bean plants by prompt chlorophyll fluorescence measurements and Artificial Neural Networks

Vladimir Aleksandrov

Institute of Plant Physiology and Genetics, Bulgarian Academy of Science
*e-mails:* aleksandrov@gbg.bg ; aleksandrov@bio21.bas.bg

## Abstract

The deficiency of macro (N, P, S, Ca, Mg and K) and micro (Zn, Cu, B, Mo, Cl, Mn and Fe) minerals has a major effect on plant development. The lack of some nutrient minerals especially of nitrogen, potassium, calcium, phosphorus and iron is a huge problem for agriculture and early warning and prevention of the problem will be very useful for agro-industry. Methods currently used to determine nutritional deficiency in plants are soil analysis, plant tissue analysis, or combined methods. But these methods are slow and expensive. In this study, a new method for determining nutrient deficiency in plants based on the prompt fluorescence of chlorophyll *a* is proposed. In this paper bean plants are grown on a complete nutrient solution (control) were compared with those grown in a medium, which lacked one of these elements - N, P, K, Ca and Fe. In this article the mineral deficiency in nutrient solution was evaluated by the stress response of the plants estimated by leaves photosynthetic activity. The photosynthetic activity was estimated by analysis of the chlorophyll fluorescence using JIP-test approach that reflects functional activity of Photosystems I and II and of electron transfer chain between them, as well as the physiological state of the photosynthetic apparatus as whole. Next the fluorescence transient recorded from plants grown in nutrient solution with deficiency of N, P, K, Ca and Iron, as an input data in Artificial Neural Network was used. This ANN was train to recognise deficiency of N, P, K, Ca and Iron in bean plants. The results obtained were of high recognition accuracy. The ANN of fluorescence transient was presented as a possible approach to identify/predict the nutrient deficiency using the fast chlorophyll fluorescence records.

## Introduction

To achieve their life cycle and physiological functions, plants need chemical elements such as N, P, K, Ca, Mg, S, Fe, Mn, Zn, Cu, Cl, B and Mo. The elements (N, P, K, Ca, Mg, S) are required in larger quantities (>1000 mg/kg dry matter) and are called macronutrients. Other elements - iron and the elements Mn, Zn, Cu, Cl, B and Mo are required in very small amounts (<100 mg/kg dry matter) and are called micronutrients[1]. The pH levels in soils also affect the availability of nutrients for plants. All minerals are available for plants in the pH range of 5.5–6.5[2].

---

1  Khan Towhid Osman, "Plant Nutrients and Soil Fertility Management," in *Soils: Principles, Properties and Management*, ed. Khan Towhid Osman (Dordrecht: Springer Netherlands, 2013), 129–59, https://doi.org/10.1007/978-94-007-5663-2_10.

2  R. E. LUCAS and J. F. DAVIS, "RELATIONSHIPS BETWEEN PH VALUES OF ORGANIC SOILS AND AVAILABILITIES OF 12 PLANT NUTRIENTS," Soil Science 92, no. 3 (1961), https://journals.lww.com/soilsci/Fulltext/1961/09000/RELATIONSHIPS_BETWEEN_PH_VALUES_OF_ORGANIC_SOILS.5.aspx.

In this paper we will investigate how deficiency of N, P, K, Ca and Fe in *Phaseolus vulgaris* will be determine with Artificial Intelligence algorithm.

**Nitrogen** is the most important mineral for plants and its deficiency is crucial for plant vitality. Nitrogen is involved in the building of amino acids and nucleic acids, is important for biochemistry of co-enzymes, photosynthetic pigments and polyamines[3]. The chloroplast proteins contain almost 75% of the nitrogen that exists in the leaves of the plants and about 27% of this utilised in Rubisco[4]. In the chloroplast the nitrogen is associated with the light harvesting apparatus: the major integral protein complexes, including photosystem I (PSI), photosystem II (PSII), electron transport chain, peripheral proteins and ATF-synthase.

Nitrogen deficiency lead to reduction of plant size, because of breakdown Rubisco capacity for $CO_2$ fixation that leads to a decrease in photosynthesis rate and inhibits plant growth[5]. In plants with nitrogen starvation also there are decrease in chlorophyll content[6]. Furthermore, PSII activity in plants with nitrogen deficiency is interrupted at each level. This lead to decrease of electron transport rate through the electron transport chain in the thylakoid membrane[7].

**Phosphorus** is another crucial macroelement for plant growth. It is involved in the composition of ATP, DNA and RNA; the phospholipids constituting the cell membranes; of the sugar-phosphate intermediates; in photosynthesis and breathing. This element is involved in virtually all metabolic processes. It plays an important role in the assimilation of carbon and nitrogen, in energy processes and lipid metabolism.

Phosphorus deficiency in plants lead to retarded growth and low shoot/root dry matter ratio. Also P-deficiency affected the development of reproductive organs and decreased number of flowers as well as the formation of fruits and seeds[8].

The deficiency of P affect carbon metabolism in the plants, because orthophosphate ($P_i$) is a major regulator of this kind of metabolic processes. Low levels of phosphorus reduce $CO_2$ assimilation and this lead to reduction of photosynthetic electron transport rate[9].

**Potassium** ($K^+$) is very important macronutrient for plant and it is involved in plant development and productivity. The potassium ion is important for photosynthesis, osmoregulation, enzyme activation, protein synthesis, ion homeostasis. The first visual symptoms of potassium deficiency is chlorosis, which then develops into necrosis. The potassium ions are not involved directly in photosynthetic metabolism, but K-deficiency strognly affect photosynthesis because limitation of potassium lead to decrease of ATP synthesis, reduction of $CO_2$ assimilation[10]. Also K-deficiency lead to reduction of photosynthesis because of low chlorophyll content, poor chloroplast

3  Frans JM Maathuis, "Physiological Functions of Mineral Macronutrients," *Physiology and Metabolism* 12, no. 3 (June 1, 2009): 250–58, https://doi.org/10.1016/j.pbi.2009.04.003.

4  M.D. Cetner et al., "Effects of Nitrogen-Deficiency on Efficiency of Light-Harvesting Apparatus in Radish," *Plant Physiology and Biochemistry* 119 (October 1, 2017): 81–92, https://doi.org/10.1016/j.plaphy.2017.08.016.

5  Meng Wei et al., "Growth and Physiological Response to Nitrogen Deficiency and Re-Supply in Leaf-Vegetable Sweetpotato (Ipomoea Batatas Lam)," *HortScience Horts* 50, no. 5 (2015): 754–58, https://doi.org/10.21273/HORTSCI.50.5.754.

6  Duli Zhao et al., "Corn (Zea Mays L.) Growth, Leaf Pigment Concentration, Photosynthesis and Leaf Hyperspectral Reflectance Properties as Affected by Nitrogen Supply," *Plant and Soil* 257, no. 1 (November 1, 2003): 205–18, https://doi.org/10.1023/A:1026233732507.

7  Cetner et al., "Effects of Nitrogen-Deficiency on Efficiency of Light-Harvesting Apparatus in Radish."

8  Bimal Sarker and JL Karmoker, "Effect of Phosphorus Deficiency on Growth and Trasport of K+, Na+, Cl-, No₃. in Lentil Seedling (Lens Culinaris Medik. Var. Barimasur-4)," *Dhaka University Journal of Biological Sciences* 20, no. 2 (January 1, 1970), https://doi.org/10.3329/dujbs.v20i2.8967.

9  Andreas Carstensen et al., "The Impacts of Phosphorus Deficiency on the Photosynthetic Electron Transport Chain," *Plant Physiology* 177, no. 1 (May 1, 2018): 271, https://doi.org/10.1104/pp.17.01624.

10  Chokri Hafsi, Ahmed Debez, and Chedly Abdelly, "Potassium Deficiency in Plants: Effects and Signaling Cascades," *Acta Physiologiae Plantarum* 36, no. 5 (May 1, 2014): 1055–70, https://doi.org/10.1007/s11738-014-1491-2.

ultrastructure, and restricted photoassimilates transport[11]. In additon potassium play a crucial role for plant resistance of pest and disease[12].

**Calcium** deficiency is asociated with poor development of all plants, leaf necrosis and deformation. Calcium is important for plant metabolism and regulate plant structure. Calcium ion plays crucial role in membrane structure and its functionality, especially for membrane permeability[13]. Ca ions are involved in the regulation of enzyme synthesis (protein kinases or phosphatases), in the synthesis of new cell walls and this is mean that Ca-deficiency will be very harmful for the plants. Calcium deficiency disrupt plant photosynthesis because Ca is a part of in $Mn_4CaO_5$ -cluster.

**Iron** deficiency causes chlorosis due to the reduced amount of chlorophyll, and leaves without Fe are smaller than normal[14]. The photosynthesis is a sensitive process for lack of the iron because this element is important for chlorophyll synthesis and it is participates in Fe–S proteins and heme proteins ferredoxin and cytochromes in photosynthetic electron transport chain. Also Fe is represented in the cytochrome b559 and like non-heme Fe in the PS II acceptor side and in the stromal part of core proteins between quinones $Q_A$ and $Q_B$[15].

Plants emit several kinds of light: prompt fluorescence (PF), delayed fluorescence (DF), thermoluminescence and phosphorescence. For aims of our research we use chlorophyll fluorescence signals - PF and DF, emitted by plants as a data for our Artificial Intelligence algorithm. We chose chlorophyll fluorescence because this techniques appear to be more sensitive then other techniques for the study of plant physiology under stress conditions[16]. The chlorophyll fluorescence is a rapid, non-destructive method for diagnostic of plant stress conditions, especially measurement and record of fluorescence induction kinetics data. At this moment there exist two methods for measurement of the prompt chlorophyll fluorescence - a PF signal produced following a pulse-amplitude-modulated excitation and a PF signal emitted during a strong continuous actinic excitation[17]. In our experiments, we used the second way to measure PF signals. The fluorescence rise during the first second of illumination from the initial (Fo) to the maximal (Fm) fluorescence value, The nomenclature of the kinetic induction curves of the fast (up to 1–2 s) Chl a fluorescence transient is OJIPS and the analysis of these curves is called JIP test, which are based on the theory of energy fluxes in biomembranes[18]. From PF signals and the indction curves are developed different parameters which are linked with the different steps and phases of the PF transient and the redox states of PSII and respectively with the efficiencies of electron transfer between PSII and PSI. This make the JIP test analysis and parameters a possible tool for study of nutrient content of the

11  Duli Zhao, D.M. Oosterhuis, and C.W. Bednarz, "Influence of Potassium Deficiency on Photosynthesis, Chlorophyll Content, and Chloroplast Ultrastructure of Cotton Plants," *Photosynthetica* 39, no. 1 (March 1, 2001): 103–9, https://doi.org/10.1023/A:1012404204910.

12  Anna Amtmann, Stephanie Troufflard, and Patrick Armengaud, "The Effect of Potassium Nutrition on Pest and Disease Resistance in Plants," *Physiologia Plantarum* 133, no. 4 (August 1, 2008): 682–91, https://doi.org/10.1111/j.1399-3054.2008.01075.x.

13  Peter K. Hepler, "Calcium: A Central Regulator of Plant Growth and Development," *The Plant Cell* 17, no. 8 (August 1, 2005): 2142, https://doi.org/10.1105/tpc.105.032508.

14  Khan Towhid Osman, "Plant Nutrients and Soil Fertility Management," in *Soils: Principles, Properties and Management*, ed. Khan Towhid Osman (Dordrecht: Springer Netherlands, 2013), 129–59, https://doi.org/10.1007/978-94-007-5663-2_10.

15  Inmaculada Yruela, "Transition Metals in Plant Photosynthesis," *Metallomics* 5, no. 9 (2013): 1090–1109, https://doi.org/10.1039/C3MT00086A.

16  Hazem M. Kalaji et al., "Experimental in Vivo Measurements of Light Emission in Plants: A Perspective Dedicated to David Walker," *Photosynthesis Research* 114, no. 2 (December 1, 2012): 69–96, https://doi.org/10.1007/s11120-012-9780-3.

17  Hazem M. Kalaji et al., "Experimental in Vivo Measurements of Light Emission in Plants: A Perspective Dedicated to David Walker," *Photosynthesis Research* 114, no. 2 (December 1, 2012): 69–96, https://doi.org/10.1007/s11120-012-9780-3.

18  Reto J. Strasser, Merope Tsimilli-Michael, and Alaka Srivastava, "Analysis of the Chlorophyll a Fluorescence Transient," in *Chlorophyll a Fluorescence: A Signature of Photosynthesis*, ed. George Christos Papageorgiou and Govindjee (Dordrecht: Springer Netherlands, 2004), 321–62, https://doi.org/10.1007/978-1-4020-3218-9_12.

plants. [19]. The There are some articles that show that there is a good correlation between fast Chl a fluorescence and nutrient deficiency. In other words, fluorescence of the chlorophyll *a* is a good indicator of the nutritional status of plants[20][21][22]. Because of the difference in induction curves emitted by plants with different nutritional status, it is possible to use data from the OJIPS curves and parameters as an Artificial Neural Network (ANN) input data. In this study we use ANN with backpropagation of errors[23]. Artificial Neural Network is one of the most important tools in modern science. In this paper we will show how ANN could be used for recognition of nutrient deficieny of plants.

The purpose of this study is to investigate how the kinetic induction curves and JIP test parameters change due to nutrient deficiency in plants and use these parameters and curves as ANN input data to determine nutritional status in plants. In this paper we design ANN based on registration of PF signals to develop a tool for recognition of missing nutrient of plant.

## Materials and methos

### *Plant material*

Bean plants (*Phaseolus vulgaris* cv. Cheren Starozagorski) was grown in in 1 dm$^3$ dark glass pots filled with a modified Hoagland nutrient solution (see tables 1, 2 and 3 for the components of solution). Solutions were supplied with oxigen by electrical pumps and replaced every 2 days. The pH of the nutrient mediums was about 5.0 for all modified solutions. The average temperature for day/night was 26/18 °C, respectively, relative humidity was 50-60%, and the photoperiod for the day/night cycle was 16/8 h. The maximum photosynthetically active radiation was about 4000 μmol (photons) m$^{-2}$ s$^{-1}$. After a week of growth in full Hoagland solution the plants were moved to stressed nutrient mediums. 14 days after the stress application (21 days after emergence) prompt chlorophyll a fluorescence (PF) measurements were done on 9 fully developed leaves for each treatment.

Table 1
Nonmodified Hoagland solution

| Compound | Molecular weight | Concentration of stock solution | Volume of stock solution per liter of final solution | Element | Final concentration of element | |
|---|---|---|---|---|---|---|
| *Macronutrients* | g mol$^{-1}$ | gL$^{-1}$ | ml | | mM | ppm |
| KNO$_3$ | 101.10 | 101.10 | 6.0 | N | 16 | 224 |
| | | | | K | 6 | 235 |
| Ca (NO$_3$)$_2$·4H$_2$O | 236.16 | 236.16 | 4.0 | Ca | 4 | 160 |
| NH$_4$H$_2$PO$_4$ | 115.08 | 115.08 | 2.0 | P | 2 | 62 |
| MgSO$_4$·7H$_2$O | 246.48 | 246.49 | 1.0 | S | 1 | 32 |

19  Strasser, Tsimilli-Michael, and Srivastava.

20  V ALEKSANDROV et al., "Deficiency of Some Nutrient Elements in Bean and Maize Plants Analyzed by Luminescent Method," *Bulgarian Journal of Agricultural Science* Supplement 1 (2014): 24–30.

21  Hazem M. Kalaji et al., "Identification of Nutrient Deficiency in Maize and Tomato Plants by in Vivo Chlorophyll a Fluorescence Measurements," *Photosynthesis Research for Sustainability* 81 (August 1, 2014): 16–25, https://doi.org/10.1016/j.plaphy.2014.03.029.

22  M.D. Cetner et al., "Effects of Nitrogen-Deficiency on Efficiency of Light-Harvesting Apparatus in Radish," *Plant Physiology and Biochemistry* 119 (October 1, 2017): 81–92, https://doi.org/10.1016/j.plaphy.2017.08.016.

23  Daniel Svozil, Vladimír Kvasnicka, and Jiří Pospichal, "Introduction to Multi-Layer Feed-Forward Neural Networks," *Chemometrics and Intelligent Laboratory Systems* 39, no. 1 (November 1, 1997): 43–62, https://doi.org/10.1016/S0169-7439(97)00061-0.

| | | | | Mg | 1 | 24 |
|---|---|---|---|---|---|---|
| *Micronutrients* | g mol$^{-1}$ | gL$^{-1}$ | ml | | µM | ppm |
| KCl | 74.55 | 1.864 | 2.0 | Cl | 50 | 1.77 |
| H$_3$BO$_3$ | 61.83 | 0.773 | 2.0 | B | 25 | 0.27 |
| MnSO$_4$·H$_2$O | 169.01 | 0.169 | 2.0 | Mn | 2.0 | 0.11 |
| ZnSO$_4$·7H$_2$O | 287.54 | 0.288 | 2.0 | Zn | 2.0 | 0.13 |
| CuSO$_4$·5H$_2$O | 249.68 | 0.062 | 2.0 | Cu | 0.5 | 0.03 |
| NaFeDTPA(10% Fe) | 468.20 | 30.0 | 0.3–1.0 | Fe | 16.1–53.7 | 1.00-3.00 |

*Source: After Epstein 1972.*

Table 2
Modified Hoagland Solution. The composition of the various culture media in mM. The concentration of minerals was achieved by addition X cm$^3$ of concentrated stock solution (1 mol per 1 dm$^3$ ) of corresponding component per 1 dm$^3$ of medium. Numbers in the brackets indicate
the pH of each nutrient solution.

| Hoagland Solution | Full pH 5.05 | (–Ca) pH 4.89 | (–K) pH 4.82 | (–N) pH 4.87 | (–P) pH 4.94 | (–Fe) pH 5.12 |
|---|---|---|---|---|---|---|
| Ca (NO$_3$).4H$_2$O | 4 | - | 4 | - | 4 | 4 |
| KNO$_3$ | 6 | 6 | - | - | 6 | 6 |
| MgSO$_4$.7H$_2$O | 2 | - | 2 | 2 | 2 | 2 |
| NH$_4$H$_2$PO$_4$ | 2 | 2 | 2 | - | - | 2 |
| Mg (NO$_3$)$_2$.6H$_2$O | - | 4 | - | - | - | - |
| MgCl$_2$.6H$_2$O | - | - | - | - | - | - |
| Na$_2$SO$_4$ | - | 2 | - | - | - | - |
| NaNO$_3$ | - | - | 6 | - | - | - |
| CaCl$_2$ | - | - | -, | 4 | - | - |
| KCl | - | - | - | 2 | - | - |
| NaH$_2$PO$_4$ | - | - | - | 2 | - | - |
| NH$_4$NO$_3$ | - | - | - | - | 1 | - |
| 1% Iron Citrate | 1 | 1 | 1 | 1 | 1 | - |
| Microelements (solution A) | 1 | 1 | 1 | - | 1 | |
| Microelements (solution B) | - | - | - | 1 | - | - |

Table 3
Salts containing micronutrients (without iron) used in modified Hoagland solution.

| Salts containing micronutrients | Quantity  (g dm$^{-3}$ H$_2$O) | |
|---|---|---|
| | Solution A | Solution  B |
| H$_3$BO$_3$ | 2.85 | 2.85 |
| MnSO$_4$.4H$_2$O | 1.10 | - |
| ZnSO$_4$.7H$_2$O | 0.28 | - |
| CuSO$_4$.5H$_2$O | 0.10 | - |
| (NH$_4$)$_6$Mo$_7$O$_{24}$.4H$_2$O | 0.02 | - |
| NaCl | 3.12 | 3.12 |
| MnCl$_2$.4H$_2$O | - | 0.93 |
| ZnCl$_2$ | - | 0.13 |
| CuCl$_2$.2H$_2$O | - | 0.07 |

*Chlorophyll fluorescence measurement*

Induction kinetics of PF were measure with using a Multifunctional Plant Efficiency Analyzer, M-PEA (Hansatech Instrument Ltd., King's Lynn, Norfolk, PE30 4NE, UK)[24]. Before measuring every plant was kept in dark at least for 30 min. Measurements were made on the abaxial surface of fully developed leaves on the middle part of chosen leaf. Measured signal were analysed by M-PEA-data analyzer version 5.4 software (this software is laboratory designed in Dept. Biophysics and Radiobiology, Sofia University by Petko Chernev, PhD).

*JIP test parameters*

These parameters are obtained from different characteristics points of photoinduced chlorophyll fluorescence transients and are useful instrument for analysis of plant photosynthetic apparatus[25][26]. The parameters that are used in this paper are described in Table 4.

Table 4
Definition of terms and formulae for calculation of the JIP-test parameters from the Chl a fluorescence transient OJIP emitted by dark-adapted leaves.

| Fluorescence parameters | Description |
|---|---|
| $F_0$ | minimal fluorescence, when all PS II RCs are open (at t = 0) |
| $F_M$ | maximal fluorescence, when all PS II RCs are closed |
| $V_J = \frac{F_J - F_0}{F_M - F_0}$ | relative variable fluorescence at the J-step |
| $\varphi_{Po} = 1 - \frac{F_0}{F_M}$ | maximum quantum yield of primary photochemistry (at t = 0) |
| $\varphi_{Eo} = (1 - \frac{F_0}{F_M})(1 - V_J)$ | quantum yield of electron transport (at t = 0) |
| $\varphi_{Ro} = (1 - \frac{F_0}{F_M})(1 - V_I)$ | quantum yield for reduction of end electron acceptors at the PSI acceptor side (RE) |
| $\psi_{Eo} = 1 - V_J$ | probability (at t = 0) that a trapped exciton moves an electron into the electron transport chain beyond $Q_A^-$ |
| $\delta_{Ro}$ | efficiency/probability With which an electron from the intersystem electron carriers moves to reduce end electron acceptors at the PSI acceptor side (RE) |
| $\gamma_{Rc} = \frac{Chl_{RC}}{Chl_{total}}$ | probability that a PSII Chl molecule ·functions as RC |
| $k_n$ is proportional to $\frac{1}{F_M}$ | Non-photochemical de-excitation constant |
| $PI_{ABS} = \frac{\gamma_{RC}}{1 - \gamma_{RC}} \cdot \frac{\phi_{Po}}{1 - \phi_{Po}} \cdot \frac{\psi_{Eo}}{1 - \psi_{Eo}}$ | performance index (potential) for energy conservation from exciton to the reduction of intersystem electron acceptors |

24 Reto J. Strasser et al., "Simultaneous in Vivo Recording of Prompt and Delayed Fluorescence and 820-Nm Reflection Changes during Drying and after Rehydration of the Resurrection Plant Haberlea Rhodopensis," *16th European Bioenergetics Conference 2010* 1797, no. 6 (June 1, 2010): 1313–26, https://doi.org/10.1016/j.bbabio.2010.03.008.

25 Strasser, Tsimilli-Michael, and Srivastava, "Analysis of the Chlorophyll a Fluorescence Transient."

26 Strasser et al., "Simultaneous in Vivo Recording of Prompt and Delayed Fluorescence and 820-Nm Reflection Changes during Drying and after Rehydration of the Resurrection Plant Haberlea Rhodopensis."

| | |
|---|---|
| $PI_{total} = PI_{ABS} \frac{\delta_{Ro}}{1-\delta_{Ro}}$ | performance index (potential) for energy conservation from exciton to the reduction of PSI end acceptors |
| $ABS/RC = \frac{1-\gamma_{RC}}{\gamma_{RC}}$ | absorption flux (of antenna Chls) per RC |
| $M_0$ | approximated initial slope (in $ms^{-1}$) of the fluorescence transient $V = f(t)$ |
| $TR_o/RC = M_0 \left(\frac{1}{V_J}\right)$ | trapping flux (leading to $Q_A$ reduction) per RC |
| $ET_o/RC$ | electron transport flux (further than $Q_A^-$) per RC |
| $RE_o/RC$ | electron flux reducing end electron acceptors at the PSI acceptor side, per RC |
| $RC/CS_o = \varphi_{Po} F_0 \left(\frac{V_J}{M_0}\right)$ | density of RCs ($Q_A^-$ reducing PSII reaction centres) |

## Statistical analysis

All experiment data were statistically analysed and the non-parametric Kruskale-Wallis one-way analysis of variance by ranks was applied.

## Artificial Neural Network

The Artificial Neural Networks are computer models based on ideas for multiple regression and classification analysis that consist of several elements operating in parallel. The functionality and capacity of the network depend on the links between the neurons that build it, and the way they are located (for details see fig. 1). For an ANN to function, it must be trained for the work it will perform. The topology of Artificial Neural Network is formed from nods (neurons) which are grouped in layers. The first layer is called input layer. The last layer is called output layer. Between them there are other layers that are called hidden layers or computational layers[27]. Depending on the structure and way of learning, there are different types of ANNs: feed-forward Artificial Neural Networks, recurrent Artificial Neural Networks, Elman and Jordan Artificial Neural Networks, long short term memory, Bi-directional Artificial Neural Networks (Bi-ANN), Self-Organizing Map (SOM), stochastic Artificial Neural Network. There exist three major types of learning: supervised learning, unsupervised learning and reinforcement learning[28].

In this study, we used feed-forward Artificial Neural Network with supervised learning. The supervised learning is machine learning algorithm that is used for training of ANN to recognize or classified data. For this purpose a training data set is used. The training data consist input and desirable output data. To be achieved complete training of ANN the input and output data have to be fit. In our work is used learning algorithm called - backpropagation of errors. This type of learning consist of two passes trough the layers of ANN - a forward pass and a backward pass. In the forward pass, an active signal are applied to input layer and the signal propagate through hidden layers to output layer. When this signal reach the output layer, it produce signal in response of input signal and than the output signal is pass back to input layer. The propagated output signal changes the input layer in such a way that the next input signal to produce an output signal with close properties to desirable output signal. This process is repeated until it reaches the desired signal[29].

27  Oludare Isaac Abiodun et al., "State-of-the-Art in Artificial Neural Network Applications: A Survey," *Heliyon* 4, no. 11 (November 1, 2018): e00938, https://doi.org/10.1016/j.heliyon.2018.e00938.

28  Andrej Krenker, "Introduction to the Artificial Neural Networks," in *Artificial Neural Networks*, ed. Janez Bešter (Rijeka: IntechOpen, 2011), Ch. 1, https://doi.org/10.5772/15751.

29  David E. Rumelhart, Geoffrey E. Hinton, and Ronald J. Williams, "Learning Representations by Back-Propagating Errors," *Nature* 323, no. 6088 (October 1, 1986): 533–36, https://doi.org/10.1038/323533a0.

The code for our ANN is written on Python. We created free-forward neural network with with hyperbolic tangent sigmoid transfer function in the hidden layers and a linear transfer function in the output layer[3031].

The input signals in our ANNs are PF induction curves and JIP test parameters recorded and calculated respectively from leguminous plants grown hydroponically in nutrient mediums under various nutrient deficiencies. The number of the input induction curves is 150 for each deficient and for control plants. The hidden neurons are 8 and input parameters are 2 - induction curves for control plants and induction curves for plants with deficiency of some nutrient element. The ANNs were trained with 600, 800 and 1000 repeats of the learning algorithm. The 3/4 of data set for each nutrient used for deficiency was training of the ANN's.

## Results

*Chlorophyll a fluorescence and JIP test*

Prompt chlorophyll fluorescence changes are considered to be sensitive indicator for nutrient deficiency in plants[323334]. The PF is measured from all leaves displayed OJIPS transients when plotted on a logarithmic time scale. The OJIPS curves in health plants have two points between O and P points. J is displayed about 2 ms and I is 30 ms after the beginning of fluorescence emitted by the chlorophyll. The OJ phase depend on light and contains information on antenna size and connectivity between PSII reaction centres[35]. The rise of transient from J to P is called thermal phase and depend on reduction of the rest of the electron transport chain[36].

In our studies we observed that the curves of induction kinetics for different deficits were changed.

30  Grady Hanrahan, *Artificial Neural Networks in Biological and Environmental Analysis*, First, Analytical Chemistry (CRC Press, 2017).

31  Simon O. Haykin, *Neural Networks: A Comprehensive Foundation*, 2nd e

32  V ALEKSANDROV et al., "Deficiency of Some Nutrient Elements in Bean and Maize Plants Analyzed by Luminescent Method," *Bulgarian Journal of Agricultural Science* Supplement 1 (2014): 24–30.

33  M.D. Cetner et al., "Effects of Nitrogen-Deficiency on Efficiency of Light-Harvesting Apparatus in Radish," *Plant Physiology and Biochemistry* 119 (October 1, 2017): 81–92, https://doi.org/10.1016/j.plaphy.2017.08.016.

34  Hazem M. Kalaji et al., "Identification of Nutrient Deficiency in Maize and Tomato Plants by in Vivo Chlorophyll a Fluorescence Measurements," *Photosynthesis Research for Sustainability* 81 (August 1, 2014): 16–25, https://doi.org/10.1016/j.plaphy.2014.03.029.

35  A. Stirbet et al., "Modeling Chlorophyll a Fluorescence Transient: Relation to Photosynthesis," *Biochemistry (Moscow)* 79, no. 4 (April 1, 2014): 291–323, https://doi.org/10.1134/S0006297914040014.

36  Gert Schansker, Szilvia Z. Tóth, and Reto J. Strasser, "Methylviologen and Dibromothymoquinone Treatments of Pea Leaves Reveal the Role of Photosystem I in the Chl a Fluorescence Rise OJIP," *Biochimica et Biophysica Acta (BBA) - Bioenergetics* 1706, no. 3 (February 17, 2005): 250–61, https://doi.org/10.1016/j.bbabio.2004.11.006.

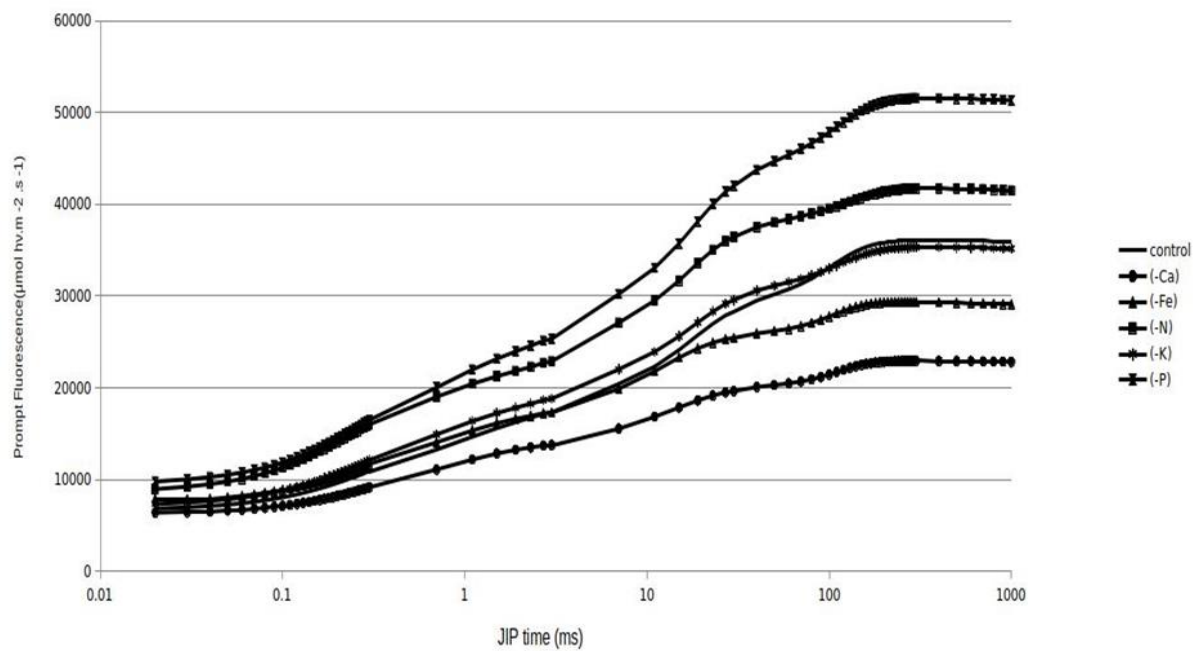

Fig. 1 Induction curves of PF, measured in Phaseolus vulgaris leaf, control and grown in Ca, N, K, P and Fe nutrient deficiencies. Fluorescence was measured by illumination of the plant with red light with an intensity of 4000 μmol hw m $^{-2}$ s $^{-1}$.

Figure 1 shows the fluorescence curves measured in Phaseolus vulgaris L. (Cheren Starozagorski), grown as a water culture in Hougland complete solution (control) or in modified Hoagland solution for nutrient deficiency. The transient induction curves of chlorophyll fluorescence are presented in logarithmic scale of time. On all induction curves the characteristic J and I phases are clearly observable. The phase J is due of accumulation of $Q_A-$. By the level of the relative fluorescence in the J phase, we can understand how the various deficiencies affect the photoinduced reduction of $Q_A$ and the subsequent oxidation of $Q_A-$ by the photosynthetic electron transport chain.

In table 5 the values of JIP test parameters for plants grow in deficiency solution are presented. Only parameters with statistically significant differences in the values between control and experimental plants will be discussed.

Table 5

Calculated JIP parameters in relative units for N, P, K, Ca and Fe deficiency *Phaseolus vulgaris* plants, normalized to respective parameter values, calculated for control plants.

Table 5a (-N)

| JIP parameters | | $V_J$ | $V_I$ | $\delta(R_0)$ | $\varphi(P_0)$ | $\varphi(E_0)$ | $\psi(E_0)$ | $\gamma(RC)$ |
|---|---|---|---|---|---|---|---|---|
| Control | | 1.00 | 1.00 | 1.00 | 1.00 | 1.00 | 1.00 | 1.00 |
| (-N) | | 1.23±0.005 | 1.14±0.04 | 0.69±0.004 | 0.98±0.08* | 0.85±0.005 | 0.86±0.005 | 0.84±0.005 |
| JIP parameters | ABS/RC | $TR_0/RC$ | $M_0$ | RC/ABS | PI(ABS) | PI(total) | $ET_0/RC$ | $RE_0/RC$ |
| (-N) | 1.32±0.003 | 1.36±0.003 | 1.69±0.004 | 0.75±0.005 | 0.53±0.002 | 0.29±0.003 | 1.16±0.04 | 0.81±0. |

Table 5b (-P)

| JIP parameters | | $V_J$ | $V_I$ | $\delta(R_0)$ | $\varphi(P_0)$ | $\varphi(E_0)$ | $\psi(E_0)$ | $\gamma(RC)$ |
|---|---|---|---|---|---|---|---|---|
| Control | | 1.00 | 1.00 | 1.00 | 1.00 | 1.00 | 1.00 | 1.00 |
| (-P) | | 1.06±0.05 | 1.06±0.05 | 0.87±0.004 | 1.02±0.08* | 0.98±0.08* | 0.97±0.07* | 0.90±0.05 |
| JIP | ABS/RC | $TR_0/RC$ | $M_0$ | RC/ABS | PI(ABS) | PI(total) | $ET_0/RC$ | $RE_0/RC$ |

| parameters | | | | | | | |
|---|---|---|---|---|---|---|---|
| (-P) | 1.19±0.04 | 1.17±0.04 | 1.23±0.005 | 0.84±0.006 | 0.82±0.006 | 0.66±0.005 | 1.13±0.04 | 0.99± |

Table 5c (-K)

| JIP parameters | $V_J$ | $V_I$ | $\delta(R_0)$ | $\varphi(P_0)$ | $\varphi(E_0)$ | $\psi(E_0)$ | $\gamma(RC)$ |
|---|---|---|---|---|---|---|---|
| Control | 1.00 | 1.00 | 1.00 | 1.00 | 1.00 | 1.00 | 1.00 |
| (-K) | 1.17±0.005 | 1.09±0.05 | 0.83±0.005 | 0.99±0.09* | 0.90±0.05 | 0.90±0.05 | 0.93±0.05 |

| JIP parameters | ABS/RC | $TR_0$/RC | $M_0$ | RC/ABS | PI(ABS) | PI(total) | $ET_0$/RC | $RE_0$/RC |
|---|---|---|---|---|---|---|---|---|
| (-K) | 1.13±0.03 | 1.14±0.03 | 1.33±0.003 | 0.88±0.007 | 0.65±0.005 | 0.52±0.002 | 1.03±0.07* | 0.85±0.0 |

Table 5d (-Ca)

| JIP parameters | $V_J$ | $V_I$ | $\delta(R_0)$ | $\varphi(P_0)$ | $\varphi(E_0)$ | $\psi(E_0)$ | $\gamma(RC)$ |
|---|---|---|---|---|---|---|---|
| Control | 1.00 | 1.00 | 1.00 | 1.00 | 1.00 | 1.00 | 1.00 |
| (-Ca) | 1.57±0.005 | 1.13±0.05 | 0.93±0.05 | 0.74±0.005 | 0.58±0.003 | 0.67±0.005 | 1.15±0.05 |

| JIP parameters | ABS/RC | $TR_0$/RC | $M_0$ | RC/ABS | PI(ABS) | PI(total) | $ET_0$/RC | $RE_0$/RC |
|---|---|---|---|---|---|---|---|---|
| (-Ca) | 0.81±0.007 | 1.31±0.007 | 2.25±0.003 | 1.40±0.006 | 0.40±0.005 | 0.36±0.008 | 0.75±0.007 | 0.71±0.007 |

Table 5e (-Fe)

| JIP parameters | $V_J$ | $V_I$ | $\delta(Ro)$ | $\varphi(P_0)$ | $\varphi(E_0)$ | $\psi(E_0)$ | $\gamma(RC)$ |
|---|---|---|---|---|---|---|---|
| Control | 1.00 | 1.00 | 1.00 | 1.00 | 1.00 | 1.00 | 1.00 |
| (-Fe) | 0.75±0.006 | 0.63±0.005 | 1.42±0.009 | 1.16±0.002 | 0.69±0.005 | 0.75±0.008 | 0.68±0.005 |

| JIP parameters | ABS/RC | $TR_0$/RC | $M_0$ | RC/ABS | PI(ABS) | PI(total) | $ET_0$/RC | $RE_0$/RC |
|---|---|---|---|---|---|---|---|---|
| (-Fe) | 0.76±0.006 | 0.80±0.006 | 0.99±0.09* | 0.73±0.006 | 0.41±0.008 | 0.30±0.009 | 0.69±0.005 | 0.55±0.003 |

deviations from the control values are unreliable at a level of significance <0.05.

*Artificial Neural Network*

JIP analyzes made for plants grown in the absence of a nutrient element indicate that the ICs of the PF and the JIP test parameters are different for each missing element. Based on these differences, we designed ANH to identify the missing nutrient in plants. To construct ANN, as input data, we used the induction curves measured in the Phaseolus vulgaris leaf as well as some JIP test parameters. The plants were grown in modified Hoagland solution at different nutrient deficiency. In our case the deficits presented in the network were: (-Fe), (-K), (-N), (-P), (-Ca) and (Con) - control plants grown in a complete nutrient medium. We have 6-component output vector of the type [1, 0, 0, 0, 0, 0], and 1 match the data for the first deficit delivered at the entrance of ANN. Once the data for the first deficit is submitted, the data for the second is given, with the output vector having the form [0, 1, 0, 0, 0, 0] and so on. We used 3/4 of network training data using the Bayesian Regularization method by training our network 600 to 1000 times. The first task was measured fluorescence signals of PF of bean plants to be used as input network data and to check whether these signals can be used to detect nutrient deficiencies in the plants. The data that was obtained during the training of the network are presented in Table 5. Using all induction curves (representing the plants with nutrient deficiency (-Fe), (-K), (-N), (-P), (-Ca) and the control plants), the optimal Network operation for 600, 800, and 1000 reps is reached at 8 hidden neurons. The increase of the number of hidden neurons does not increase network accuracy. On the other hand, the increase the repeatability of the training (epochs) after 800 repetitions in practice not increases the accuracy of the network and for this reason we have accepted that the optimal number of reps is 800.

Table 6

Input and output data of ANN trained to detect nutritional deficiencies in Phaseolus vulgaris. The number of trains of the network (epochs) varies from 600 to 1000. The number of hidden neurons varies from 2 to 10 and the input parameters are 6, which corresponds to measurements of the fluorescence signals of PF for plants grown in environments with different nutrient deficiencies (5 variants) + signal from control plants.

| Input data and repetitions | Input parameters | Hidden Neurons | Wrong answers (%) | | |
|---|---|---|---|---|---|
| | | | Training | Test | Total |
| *Phaseolus vulgaris* PF (600) | 6 | 2 | 43.6 | 43.8 | 43.7 |
| *Phaseolus vulgaris* PF (800) | 6 | 2 | 43.2 | 43.6 | 43.5 |
| *Phaseolus vulgaris* PF (1000) | 6 | 2 | 43.3 | 43.6 | 43.4 |
| *Phaseolus vulgaris* PF (600) | 6 | 4 | 25.7 | 35.8 | 28.2 |
| *Phaseolus vulgaris* PF (800) | 6 | 4 | 25.5 | 33.9 | 27.6 |
| *Phaseolus vulgaris* PF (1000) | 6 | 4 | 25.3 | 32.8 | 27.0 |
| *Phaseolus vulgaris* PF (600) | 6 | 8 | 25.9 | 29.6 | 26.8 |
| *Phaseolus vulgaris* PF (800) | 6 | 8 | 25.2 | 28.4 | 26.0 |
| *Phaseolus vulgaris* PF (1000) | 6 | 8 | 25.3 | 28.6 | 26.5 |
| *Phaseolus vulgaris* PF (600) | 6 | 10 | 25.5 | 30.2 | 26.7 |
| *Phaseolus vulgaris* PF (800) | 6 | 10 | 25.2 | 29.8 | 26.3 |
| *Phaseolus vulgaris* PF (1000) | 6 | 10 | 24.9 | 29.6 | 26.2 |

In Table 7 are represented the results obtained when the network is trained to detect only one nutrient deficiency. As input data we use two elements - the data on one macro element deficiency and the data on control plants.

Table 7

Input and output data of ANN trained to detect nutritional deficiencies in Phaseolus vulgaris. For the input parameters the data obtained by measuring the fluorescence signals of PF are used simultaneously. The number of network exercises (epochs) is 800 iterations. The number of hidden neurons is 8 and the input parameters are 2, which corresponds to a signal measured from a plant grown in a nutrient mineral deficiency + signal from control plants.

| Input data and repetitions | Input parameters | Hidden Neurons | Wrong answers (%) | | |
|---|---|---|---|---|---|
| | | | Training | Test | Total |
| *Phaseolus vulgaris* (800) (–Fe) – Controls | 2 | 8 | 3.1 | 4.0 | 3.5 |
| *Phaseolus vulgaris* (800) (–K) – Controls | 2 | 8 | 0.2 | 1.8 | 0.6 |
| *Phaseolus vulgaris* (800) (–N) – Controls | 2 | 8 | 2.3 | 5.3 | 3.0 |
| *Phaseolus vulgaris* (800) (–P) – Controls | 2 | 8 | 0 | 0 | 0 |
| *Phaseolus vulgaris* (800) (–Ca) – Controls | 2 | 8 | 1.1 | 5.2 | 2.4 |
| *Phaseolus vulgaris* (800) - ALL | 6 | 8 | 51.6 | 52.9 | 51.9 |

# Discussion

For the normal physiological state of the plant the availability of micro and macro-elements is very important. In this study we measured chlorophyll a flourescence transient to analyse the changes in light phase of photosynthesis in nutrient-deficient bean plants. The plants were grown hydroponically to determine possible effect of macronutrients (N, P, K and Ca) and micronutrient (Fe) deficiency on the Electron Transport Chain in the chloroplasts. Nutrient deficiency induced changes in chlorophyll a fluorescence induction curve as well as in JIP test parameters. Because of these changes in fluorescence induction curve and in JIP test parameters we could use both of them for nutrient deficiency recognition.

**Nitrogen deficiency.** The value of Fo for plants with nitrogen deficiency is higher than the value of Fo for control plants. The higher initial fluorescence value, measured in the absence of nitrogen at the plant is proof of the lower efficiency of transmitting of the excitation energy from the Light Harvesting Complex (LHCII)to the reaction centers of PSII[37]. On the other hand, plants subject to nitrogen deficiency have a higher value for $F_M$ compared to the control plants. The $V_j$ and $V_i$ values of the control plants are less than those of plants with nitrogen deficiency and this means that PSI oxidizes stronger the plaquequinone pool in plants that develop in the absence of nitrogen. The lower value of $\psi(Eo)$ for nitrogen-stressed plants shows that electron transport after primary quinone is much less likely. The lack of nitrogen in the plants leads to a reduction in electron transport by ETC, reflected by the parameter $\varphi(Eo)$. The value of parameter $\gamma(RC)$ for stressed plants is lower than the value of this parameter for control plants. This means that the relative amount of chlorophyll molecules acting as RCs in plants grown in nitrogen deficiency is less than in control plants. The fact that the value of REo/RC is lower in plants grown under nitrogen deficiency compared to control plants indicates that much less electrons manage to reduce the last acceptors of PSI. The two performance indices PI(ABS) and PI(total) have very low values for stressed plants compared to unstressed. This shows that, in general, the lack of nitrogen has a strong negative effect on the photosynthetic apparatus. The higher value of the TRo/RC parameter for nitrogen deficiency plants shows that they capture more energy in the RC than the control plants. The value of the N parameter for the plants with nitrogen deficiency is lower than the value of this parameter for control plants. Therefore, fewer electrons are required for the complete recovery of acceptors after $Q_A$.

**Phosphorus deficiency.** The value of minimal fluorescence signal (Fo) in plants with phosphorus deficiency is higher then the value of Fo in control plants as well as Fm. The parameters $\delta(Ro)$ and $\gamma(RC)$ are lower values then the values of the control plants. The first of these parameters gives information about the ability of the intermediate carriers to reduce the last acceptors of PS I. The lower value for $\gamma(RC)$ indicates that the plants with a deficiency of phosphorus have a relatively smaller number of RC relative to plants grown in normal environment. In the case of phosphorus deficiency, the value for N is lower than in the control plants. The values of ABS/RC and TRo/RC for plants with phosphorus deficiency are higher than in unstressed plants. The ETo/RC parameter, which gives information about the flow of electrons after $Q_A$ is greater in plants with phosphorus deficiency than in unstressed ones. This is an indication that a greater number of electrons are able to pass ETC after the primary quinone acceptor in PSII. On the other hand, the REo/RC parameter is not substantially altered, indicating that approximately the same number of electrons in both stressed and unstressed plants reach to the final acceptors of PSI. This means that for stressed plants the losses of energy are mainly observed at intermediate carriers in ETC. One reason for this may be the water-water cycle[38].

37 Michel Havaux, Hubert Greppin, and Reto J. Strasser, "Functioning of Photosystems I and II in Pea Leaves Exposed to Heat Stress in the Presence or Absence of Light," *Planta* 186, no. 1 (December 1, 1991): 88–98, https://doi.org/10.1007/BF00201502.

38 X. -Y. Weng et al., "Water-Water Cycle Involved in Dissipation of Excess Photon Energy in Phosphorus Deficient Rice Leaves," *Biologia Plantarum* 52, no. 2 (June 1, 2008): 307, https://doi.org/10.1007/s10535-008-0064-x.

**Potassium deficiency.** The ICs of control plants and those with K-deficiency are distinguished slightly from each other, indicating that the lack of K does not significantly affect the photosynthetic apparatus. This may be due to partial replacement of missing potassium ions with sodium ions in processes related to photosynthesis. Significant differences exist only with two parameters: $V_J$ and $\delta(R_o)$. The first parameter tells us that in plants lacking potassium there is a relatively greater number of closed reaction centers at the J level of IC compared to control plants. The lower values for the parameter $\delta(R_o)$ show that the probability of reducing the last acceptors of PS I is lower for plants with potassium deficiency. The JIP parameters ABS/RC, TRo/RC, Mo, RC/ABS, PI(ABS), PI(total) and REo/RC are altered due to potassium deficiency. Low values for PI(ABS) and PI(total) indicate that lack of potassium leads to changes in ETC and lowering the photosynthetic activity of the plants.

**Calcium deficiency.** For plants growing in a calcium-free environment, fluorescence is much less intense than that of the control plants. This is evident from the large difference between the value of the parameter $F_V$ of the stressed and unstressed plants. Lack of calcium causes changes in almost all JIP parameters. The PI(ABS) and PI(total) parameters have very low values for stressed plants relative to unstressed. JIP analysis shows that calcium deficiency has an extremely strong impact on the photosynthetic apparatus and affects almost all of its components.

**Iron deficiency.** The higher value for Fo indicates that the light harvesting complex of the stressed plants is less effective than the light harvesting complex of the control plants. For bean plants grown in iron deficient environments, the parameters $\varphi(Po)$ and $\delta(R_o)$ have higher values than in the control plants. The first of the two parameters gives information that in the stressed plants the transfer of electrons from RC to $Q_A$ is more likely. The second parameter indicates that the probability, with which electron reduces the last acceptors of PS I, is greater for plants grown in Iron free environments. $\varphi(Eo)$ and $\psi(Eo)$ provide information that the probability of transfer of electrons after the primary quinone acceptor is much lower for stressed plants. The reason for this is probably due to the lack of intermediate non-heme iron between $Q_A$ and $Q_B$. The lack of iron in plants has a negative impact on the whole photosynthetic apparatus.

It is clear that deficiency of some nutrients lead to difference in JIP test parameters and fluorescence induction curves in plants. This gave us reason to use artificial neural network for fast and accuracy recognition of nutrient deficiency in bean plants. The network which we used in our work was ANN with backpropagation of error.

We used ¾ of network training data using the *Bayesian Regularization* method and we trained the network from 600 to 1000 times.

The first task was to use the measured fluorescence signals of PF of bean plants to be used as incoming network data and to check if it can be used to detect nutrient deficiencies. Firts we trained the network 600 times. We first trained the network with 600 iterations by changing the number of input parameters (deficiencies) and the number of hidden neurons.The number of induction curves is 648. The data obtained during the training of the network are presented in Table 6. Using all parameters (representing plants with 5 types of tested deficits: (-Fe), (-K), (-N), (-P), (-Ca) and control plants) optimal network operation for 600, 800 and 1000 reps are reached in 8 hidden neurons. Increasing the number of hidden neurons does not increase the accuracy of the network. On the other hand, increasing repetition of training (epochs) after 800 iterations does not actually increase the accuracy of the network and therefore we have accepted that the optimal number of iterations is 800.

The next task was to use as an input network data only two signals - one for the signals measured by the deficiency plants and signals measured by the control plants. The results are presented in Table 7. From Table 6 and Table 7 it is evident that the trained network for only two parameters gives a much smaller error compare to network trained to detect all deficiencies at once.

The network gives the biggest training error to recognize iron and nitrogen deficiency. The network trained to detect phosphorus deficiency does not produce any erroneous results. On the other hand, when we submitted the total number of data on all analyzed options: control and deficient, the error increased to 52%.

This gives us reason to assume that the appropriate strategy for recognizing nutrient deficiencies is for the network to be trained to recognize each deficiency individually.

The record of OJIP transient in our experiments and the JIP test allowed us to quantify photosynthetic parameters that were significant for the evaluation of the photosynthetic apparatus of the investigated plants subjected of nutrient deficiency stress. From our results it is clear that the same photosynthetic parameters calculated for plants subject to a different nutritional deficiency have different values. These results are important, because they shown that some photosynthetic partameters are sensitive to nutrient deficiency and could be used as a fluorescence phenotype marker.

Applying the AI to OJIP transient data allows us to recognize which nutrients are missing in plants. This approach allows to be developed fast and high accurate method for plants monitoring *in vivo* conditions.

## Conclusion

Deficiency of all analysed elements changed the physiological state of bean plants that was displayed in modifications of the chlorophyll fluorescence transients. The effects of the lack of these elements included the impairments in electron transport chain in both donor and acceptor sides of PSII and of PSI. The Artificial Newral Network with backpropagation was applied to recognize nutrient deficiency on the basis of chlorophyll fluorescence data. Our results suggest that the Artificial Newral Network approach for early recognition of nutrient deficiency based on chlorophyll fluorescence data is very useful and powerful tool.